\documentclass[a4paper,12pt]{article}        
\usepackage{hyperref}
\usepackage{graphicx}
\usepackage[english]{babel}
\def\d{\delta}          \def\eps{{\epsilon}}     
                    \def\l{\lambda}
\def\m{\mu}             \def\n{\nu}              
                       
\def\vphi{\varphi}      \def\om{\omega}
          \def\D{\Delta}           \def\L{\Lambda}

\def\be{\begin{equation}}            \def\ee{\end{equation}}
\def\ba#1{\begin{array}{#1}}         \def\ea{\end{array}}

\def\fr#1#2{\textstyle\frac{#1}{#2}}

\begin{document}

\title{\Large\bf  Extrasolar scale change in   
 Newton's Law \\ from 5D `plain' R$^2$-gravity
 (on `very thick brane')}
\author{I L Zhogin (SSRC, Novosibirsk)\footnote{
 E-mail: zhogin at inp.nsk.su; \
 \href{http://zhogin.narod.ru}{http://zhogin.narod.ru }
} }
\date{}
 \maketitle

\begin{abstract}
Galactic rotation curves and lack of direct
observations of Dark Matter may indicate
that General Relativity is not valid
(on  galactic scale) and should be replaced
with another theory.

There is the only variant of Absolute Parallelism
 which  solutions
are free of arising singularities, if
D=5 (there is no room for changes).
This variant 
 does not have a Lagrangian, nor match GR:
 an equation of `plain' R$^2$-gravity
 (ie without R-term)
 is in sight instead.

Arranging an
 expanding O$_4$-symmetrical
solution as the basis of 5D cosmological model,
and probing a \emph{universal function} of
mass distribution (along very-very long the extra dimension)
to place into bi-Laplace equation
(R$^2$ gravity),
one can derive the Law of Gravitation:
 $\fr1{r^2}$ transforms to $\fr1{r}$ with distance (not
with acceleration).

\end{abstract}

 \section{\large Introduction}
 Being a `close relative' of
 General Relativity (GR),
 Absolute Parallelism (AP) has many interesting features: larger
symmetry group of equations; field irreducibility with respect to
this group; vast list of compatible second order equations
(discovered by Einstein and Mayer \cite{eima}) not
restricted to Lagrangian ones.

There is the only variant of Absolute Parallelism
 which  solutions
are free of arising singularities, if
D=5 (there is no room for changes; this variant of AP
does not have a Lagrangian, nor match GR);
in this case AP has topological features of nonlinear
sigma-model.

In order to give clear presentation and full picture of the
theory' scope, many items should be sketched: instability of
trivial solution and expanding O$_4$-symmetrical ones; tensor
$T_{\mu\nu}$ (positive energy, but only three polarizations of
15 carry (and angular) momentum; how to quantize such a stuff ?)
 and PN-effects; topological
classification of symmetric 5D field configurations (alighting
on evident parallels with Standard Model' particle
combinatorics) and `quantum phenomenology on expanding classical
background' (coexistence); `plain' R$^2$-gravity on very thick
brane and change in the Newton's Law: $\fr1{r^2}$ goes to
$\fr1{r}$ with distance (not with acceleration -- as it is in
MOND \cite{mond}).

 At last, an experiment with single photon
interference is discussed as the other way to observe very-very long
(and very undeveloped) the extra dimension.
\section{\large Unique 5D equation of AP
 (free of singularities in solutions)}

There is one unique variant of AP (non-Lagrangian, with the unique
$D$; $D$=5) which solutions of general position seem to be free
of arising singularities.
 The formal integrability test
 \cite{pommaret} can be extended to the cases of degeneration of
either co-frame matrix,
$h^a{}_\mu$, (co-singularities) or
contra-variant frame (or contra-frame density of some weight),
serving as the local and covariant (no coordinate
choice) test for singularities of solutions. In AP this test
singles out the next equation
(and D=5, see  \cite{tc};
$\eta_{ab}=\mbox{diag}(-1,1,\ldots,1)$, then
$h=\det h^a{}_\mu=\sqrt{-g}$):
 \begin{equation} \label{ue}
 {\bf E}_{a\mu}=L_{a\mu\nu;\nu}- \fr13 (f_{a\mu}
 +L_{a\mu \nu }\Phi _{\nu })=0\, ,
\end{equation}
 where (see \cite{tc} for more detailed introduction to AP and
 explanation of notations used)
 \[ L_{a\mu \nu }=L_{a[\mu \nu]}=
\Lambda_{a\mu \nu }-S_{a\mu \nu }-\fr23 h_{a[\mu }\Phi_{\nu]},
\]
\begin{equation}\label{defin}
\Lambda_{a\mu \nu }=2h_{a[\mu,\nu]}, \
S_{\mu \nu \l}=3\L_{[\mu \nu \l]}, \
\Phi_\mu=\L_{aa \mu}, \
f_{\mu\nu}=2\Phi_{[\mu,\nu]}=2\Phi_{[\mu;\nu]}  .
\end{equation}
Coma "," and semicolon ";" denote partial
derivative and usual covariant differentiation with symmetric
Levi-Civita connection, respectively.

One should retain the
identities [which follow from the definitions (\ref{defin})]:
\begin{equation}\label{ident}
 \Lambda_{a[\mu\nu;\lambda]} \equiv 0\,,
  \ \  h_{a\l}\Lambda_{abc;\l}\equiv f_{cb}\,
(= f_{\m\n}h_c^\m h_b^\n), \ f_{[\m\n;\l]}\equiv0.
\end{equation}

The equation ${\bf E}_{a\mu;\mu}=0$ gives
 `Maxwell-like equation' (we prefer to omit
 $g^{\m\n} \ (\eta^{ab})$
in contractions that not to keep redundant information
-- when covariant differentiation is in use only):
\begin{equation}\label{max}
(f_{a\mu}
 +L_{a\mu \nu }\Phi _{\nu })_{;\mu}=0, \mbox{ or \ }
 f_{\mu\nu;\nu}=(S_{\mu \nu\l }\Phi _{\l })_{;\n} \ \
(= -\fr1 2 S_{\mu \nu\l }f_{\n\l}, \mbox{ see below})
\, .
\end{equation}
Actually the Eq.~(\ref{max}) follows from the symmetric part
of equation, ${\bf E}_{(ab)}$,
because skewsymmetric one gives just the identity:
 \[
2{\bf E}_{[\nu\mu]}=S_{\mu\nu\l;\l}=0, \
{\bf E}_{[\mu\nu];\nu}\equiv 0; \]
note also that the trace part
 becomes irregular (the principal
 derivatives vanish) if $D=4$ (this number of
 dimension is forbidden, and the next number,
 $D=5$, is the most preferable):
\[{\bf E}_{\m\m}={\bf E}_{a\m}h_b^\m\eta^{ab}
=\fr{4-D}3 \Phi_{\m;\mu}+ (\L^2)=0.
  \]

The system (\ref{ue}) remains compatible under adding
$f_{\m\n}=0$, see (\ref{max});  this is not the case for
another covariant, $S, \Phi$, or (some irreducible part of the)
Riemannian curvature,
which relates to $\L$ as usually:
\[ R_{a\mu\nu\lambda}=
2h_{a\mu;[\nu;\lambda]}; \
h_{a\m}h_{a\nu;\l}=\fr12 S_{\m\n\l}-\L_{\l\m\n}.\]
 \section{\large Tensor
  $T_{\mu\nu}$ (despite Lagrangian absence) and PN-effects}
 One might rearrange ${\bf E}_{(\mu\nu)}{\,=\,}0$ that to pick out
 (into LHS) the Einstein tensor,
 $G_{\m\n}{\,=\,}R_{\m\n}-\fr12g_{\m\n}R$, but the rest terms are not
 proper energy-momentum tensor: they contain linear terms
 $\Phi_{(\m;\nu)}$ (no positive energy (\,!);  another
 presentation of `Maxwell equation' (\ref{max}) is possible instead
 -- as divergence of symmetrical tensor).

 However, the prolonged equation
${\bf E}_{(\mu\nu);\l;\l}=0$ can be written as `plain' (no R-term)
R$^2$-gravity:
\be \label{tmunu}{
(-h^{-1}\, \delta(h\,R_{\mu\nu}G^{\mu\nu})/\delta g_{\mu\nu}
{=})\ G_{\mu \nu ;\lambda ;\lambda }+ G_{\epsilon \tau}
(2R_{\epsilon \mu \tau \nu } - \fr12g_{\mu \nu }R_{\epsilon \tau })
=T_{\mu\nu} (\Lambda '^{2},\dots), \ T_{\mu\nu;\nu}=0;
}\ee
up to quadratic terms,
\[T_{\mu\nu}=
\fr29(\fr14g_{\m\n}f^2-f_{\m\l}f_{\n\l})
+A_{\m\eps\n\tau}(\L^2){}_{;(\eps;\tau)}+
(\L^2\Lambda ',\L^4);
\]
tensor $A$ has symmetries
of Riemann tensor, so the term $A''$ adds nothing to momentum
and angular momentum.

It is worth noting that:

(a) the theory does not match GR, but
shows `plain' $R^2$-gravity (sure, (\ref{tmunu}) does not
contain all the theory);

(b) only $f$-component
(three transverse polarizations in D=5)
carries D-momentum and angular
momentum (`powerful' waves); other 12 polarizations
are `powerless', or `weightless'
(this is a very unusual feature -- impossible
 in the Lagrangian tradition; how to quantize ?
let us not to try this, leaving the theory `as is');

(c) $f$-component feels only metric and $S$-field
(`contorsion', not `torsion' $\L$ -- to
label somehow), see (\ref{max}),
but $S$ has effect only on polarization of $f$:
$S_{[\m\n\l]}$ does not
enter eikonal equation, and $f$ moves along usual Riemannian
geodesic (if background has $f$=0);
one may think that all `quantum fields'
(phenomenological quantized fields accounting for
topological (quasi)charges and
carrying some `power';
 see further) inherit this property;

(d) the trace $T_{\m\m}=\fr1{18}f_{\m\n}f_{\m\n}$ can be
non-zero if  $f^2\neq0$ and this seemingly depends on $S$-component
[which enters the current in (\ref{max})]; in other words,
`mass distribution' is to depend on distribution of $f$-
and $S$-component;

(e) it should be stressed and underlined that the
$f$-component is not usual (quantum) EM-field --
just important covariant responsible for energy-momentum
(suffice it to say that there is no gradient invariance
for $f$).

 \section{{\large Linear domain: instability of trivial solution
 (with powerless  waves)
  }}
Another strange feature is the instability of trivial solution:
 some `powerless' polarizations grow linearly with time
 in presence of
 `powerful' $f$-polarizations. Really, from the linearized
 Eq.~(\ref{ue}) and the identity (\ref{ident}) one can write
 (the following equations should be understood as linearized):
 \[ {
 \Phi_{a,a}=0 \ (D\neq 4), \
 3\Lambda_{abd,d}= \Phi_{a,b}-2\Phi_{b,a},
 \ \Lambda_{a[bc,d],d}\equiv0\, \ \Rightarrow \
 3\Lambda_{abc,dd}=-2 f_{bc,a}\, . }\]
 The last `D`Alembert equation' has the `source' in its right
 hand side. Some components of $\Lambda$
 (most symmetrical irreducible parts)
 do not grow (as well as curvature), because
 (again, linearized equations are implied below)
 \[{ S_{abc,dd}=0, \ \Phi_{a,dd}=0, \ \,
   f_{ab,dd}=0, \ R_{abcd,ee}=0, }\]
  but the least symmetrical components of the tensor $\Lambda$
  do grow up with time (due to terms
   \ $\sim t\, e^{-i \om t}$; three growing polarizations
   which are `imponderable', or powerless)
  if  the `ponderable' waves (three $f$-polarizations)
  do not vanish (and this should be the case for
  solutions of `general position').
 \section{{\large Expanding
 O$_4$-symmetrical (single wave) solutions and cosmology }}
The unique symmetry of AP equations gives scope for symmetrical
 solutions. In contrast to GR, this variant of AP has
non-stationary spherically ($O_4$-)
 symmetric solutions.
The $O_4$-symmetric frame field  can be
generally written as follows \cite{tc}:
\begin{equation}  \label{spsy}{
 h^{a}{}_{\mu }(t,x^i)=
 \pmatrix{a&bn_{i}  \cr
 cn_{i} & en_{i}n_{j}+d\Delta _{ij} };
\ \ i,j=(1,2,3,4), \ n_i=\frac{x^i}{r}. }
\end{equation}
 Here $a,\ldots,e$ are functions of time, $t=x^0$, and radius
 $r$, $\Delta_{ij}=\delta_{ij}-n_i n_j, \ r^2=x^i x^i$.
 As functions of radius, $b,c$ are odd,
  while the others are even;
  other boundary conditions: $e=d$ at $r=0$,
 and $h^a{}_\mu\to \delta^{\,a}_\mu$ as $r\to \infty$.
Placing in (\ref{spsy})
$b=0, e=d$ (the other interesting choice is $b{=}c{=}0$)
 and making integrations one can arrive to the next system
 (resembling dynamics of Chaplygin gas; dot and prime
 denote derivation on time and radius, resp.;
  $A=a/e=e^{1/2},\ ~B=-c/e $):
\begin{equation}\label{gas}
A^{\cdot}=AB^\prime -BA^\prime +3AB/r\ ,
\ \ B^{\,\cdot }=AA^\prime
-BB^\prime -2B^{2}/r~.
\end{equation}
This system (does not suffer of gradient catastrophe and) has
non-stationary solutions; a single-wave solution of proper
`amplitude' might serve as a suitable cosmological (expanding)
background.

  The condition $f_{\mu\nu}{=}0$ is a must for solutions with such
 a high symmetry (as well as
 $S_{\mu\nu\l}{=}0$); so, these $O_4$-solutions
 carry no energy, that is, weight nothing
 (some lack of \emph{gravity} !
in this theory the universe
 expansion  seemingly has little common with
 \emph{gravity},  GR
 and its dark energy \cite{carroll}).

 More realistic cosmological model might look like a single
 $O_4$-wave
 (or a sequence of such waves) moving along the radius and being
 filled with chaos, or  stochastic waves, both
 powerful (\emph{weak}, $\Delta h\ll1$) and
  powerless ($\Delta h<1$, but intense enough that
  to lead to non-linear
  fluctuations with $\Delta h\sim1$),
  which form  statistical ensemble(s)
 having a few characteristic parameters
 (after `thermalization').
  The development and examination of stability
 of such a model is an interesting problem.
 The metric variation in  cosmological $O_4$-wave
 can serve as
 a time-dependent `shallow dielectric guide' for that weak
noise waves. The ponderable  waves
(which slightly `decelerate' the $O_4$-wave) should
have wave-vectors almost
tangent to the $S^3$-sphere of wave-front that to be trapped
inside this (`shallow') wave-guide; the imponderable waves
can grow up, and partly escape from the wave-guide,
 and their wave-vectors
can be less tangent to the $S^3$-sphere.

The waveguide thickness can be small for an observer in
the center of $O_4$-symmetry, but in co-moving coordinates
it can be very large (due to relativistic effect), however
still small with respect to the radius of sphere,
$L\ll R$. It seems that the radial dimension has to be
 very `undeveloped'; that is, there are no other characteristic
scales, smaller than $L$, along this extra-dimension.

 \section{{\large Non-linear domain: topological charges and
quasi-charges }}
Let AP-space  is of
 trivial topology: no worm-holes, no compactified space
 dimensions, no singularities.
One can continuously
deform frame field $h(x)$
to a
field of rotation matrices
(metric can be diagonalized and
 `square-rooted')
$
h^{a}{}_{\mu}(x)  \to s^{a}{}_{\mu}(x)\in SO(1,d);  \
m{=}D{-}1 $.
Further deformation can remove boosts too,
and so, for
any space-like (Cauchy) surface, this gives a (pointed) map,
\[ s\,: \ {\bf R}^m\cup \infty=S^m \to SO_m;
\ \infty\mapsto 1^m\in SO_m.\]
The set of such maps consists of homotopy classes
forming the  group
of topological charge, $\Pi(m)$:
\begin{equation}\label{pim}{
\Pi(m)=\pi_m(SO_m); \ \ \Pi(3)=Z, \ \Pi(4)=Z_2+Z_2.}
\end{equation}
Here $Z$ is the infinite cyclic group, and
$Z_2$ is the cyclic group of order two.

It is  important that deformation to $s$-field
can  keep  symmetry of field configuration.
Definition: localized field
(pointed map)
 $  s(x) :\  {\bf R}^{m} \to SO(m),\   s (\infty)=1^{m},$
is $G$-symmetric if, in some coordinates,
\begin{equation}
\label{gsi} {
  s(\sigma x)= \sigma s (x)\sigma^{-1}\
 \ \forall \ \sigma \in G \subset O(m)\, .}
\end{equation}
 The set of such fields ${\cal C}^{(m)}_G $ generally consists of
 separate, disconnected components -- homotopy classes forming
the `topological quasi-charge
 group' denoted here as  $\Pi (G;m)
\equiv\pi_{0}({\cal C}^{(m)}_G)$.
These QC-groups  classify symmetrical localized
configurations of frame field.
 Since field equation does not break symmetry,
 quasi-charge conserves;
if symmetry is not exact (because of distant regions),
quasi-charge is not exactly conserving value, and quasi-particle
(of zero topological charge) can
annihilate (or be created)
 during colliding with another quasi-particle.

 The other problem. Let $G1\supset G2$, such that there is a
 mapping (embedding)
 $  i:\ {\cal C}^{(m)}_{G1} \to {\cal C}^{(m)}_{G2}$,
which induces the homomorphism of QC-groups:
$     i_{*} :  \   \Pi(G1;m) \to  \Pi(G2;m) $,
so one has to describe this morphism.

Let us consider the simple (discreet) symmetry group  $P_1$
with a plane of reflection symmetry:
 $$P_1 =\{1,p_{(1)}\},
\mbox{ \ where } p_{(1)}=
\mbox{diag}(-1,1,\ldots,1)=p_{(1)}^{-1}.$$
 It is necessary to set field $s (x)$ on the half-space $\frac1
2\,{\bf R}^m=\{x^1\geq 0\}$, with additional condition imposed on the
surface ${\bf R}^{m-1}=\{x_1=0\}$
(stationary points of $P_1$ group) where $s $ has
to commute with the symmetry [see (\ref{gsi})]:
\[ {
p_{(1)}x=x\ \Rightarrow\ s(x)=p_{(1)}s p_{(1)}
\ \Rightarrow\ s\in 1 \times SO_{m-1}. }\]
 Hence, accounting for the localization requirement,
 we have a diad map
 (relative spheroid; here $D^m$ is an $m$-ball and
 $S^{m-1}$ its surface)
$(D^m;S^{m-1}) \to  (SO_{m}; SO_{m-1})$,
and topological classification of such maps leads to
the relative (or diad) homotopy group (\cite{du};
the last equality below follows due to fibration
$SO_{m}/SO_{m-1}=S^{m-1}$):
\[{ \Pi(P_1;m)=\pi_{m}(SO_{m};SO_{m-1})
=\pi_{m}(S^{m-1}). }\]

Similar considerations (of group orbits
and stationary points) lead to the following result:
\[{ \Pi(O_{l};m)=\pi_{m-l+1}(SO_{m-l+1};SO_{m-l})
=\pi_{m-l+1}(S^{m-l}).} \]
If $l>3$, there is the equality: $\Pi(SO_{l};m)=\Pi(O_{l};m)$,
while for $l=2,3$ one can find:
\[{ 
\Pi(SO_{3};m) = \pi_{m-2}(SO_{2}\times SO_{m-2};SO_{m-3})=
\pi_{m-2}(S^1\times S^{m-3}),
}\]
\[{ 
\Pi(SO_{2};m) = \pi_{m-1}(SO_{m};SO_{m-2}\times SO_{2})=
\pi_{m-1}(RG_+(m,2)).
}\]
\\[-11mm]

The set of quaternions  with absolute value one,
$\textbf{H}_1=\{\textbf{f},\ |\textbf{f}|=1\}$, forms a
group under quaternion multiplication,
 $\textbf{H}_1\cong SU_2=S^3$, and
any $s \in SO_{4}$
can be represented as a pair of such quaternions
\cite{du},
$({\bf f},\,{\bf g})\in
S^{3}_{(l)}\times S^{3}_{(r)},\
|{\bf f}|=|{\bf g}|=1 $:
\[ { x^*=sx \ \Leftrightarrow \ {\bf x}^*=
{\bf f\,x\,g}^{-1}={\bf f\,x\,\bar{g}} \,
; \ \ |{\bf x}|=|{\bf x^*}|. }\]
The pairs (\textbf{f},\,\textbf{g}) and
(--\textbf{f},\,--\textbf{g}) correspond to
the same rotation $s$, that is,
$SO_{4}=S^{3}_{(l)}\times S^{3}_{(r)}/\pm $.

Note that the symmetry condition (\ref{gsi}) also splits
into two parts:
\begin{equation}
\label{ab} {
 {\textbf{f}\/(\textbf{a}\/\textbf{x}\/\textbf{b}}
 {}^{-1})=\textbf{a}\/\textbf{f}(\textbf{x})\/\textbf{a} {}^{-1},
 \ \textbf{g}\/(\textbf{a}\/\textbf{x}\/\textbf{b}{}^{-1})
 =\textbf{b}\/\textbf{g}(\textbf{x})\/\textbf{b}{}^{-1} \ \forall
(\textbf{a},\/\textbf{b})\/ \in G \subset SO_{4}. }
\end{equation}

 \section{{\large Example of $SO_2$-symmetric
 quaternion field}}
Let's consider an example of $SO_2\{2,3\}-$symmetric \textbf{f}--field
 configuration (\textbf{g}=1), which carries both charge and
$SO_2$-quasi-charge (left, of course),
 $\textbf{f}(\textbf{x}){:} \
 {\bf H}={\bf R}^4 \to {\bf H}_1; \ \textbf{f}(\infty)=1$.
The symmetry condition (\ref{ab}) reads
\begin{equation}\label{f-so2} {
 \textbf{f}(e^{{\bf i}\,\phi/2}\textbf{x}e^{-{\bf i}\,\phi/2})
   =e^{{\bf i}\,\phi/2} \textbf{f}(\textbf{x}) e^{-{\bf i}\,\phi/2}.}
\end{equation}
We'll switch to `double-axial' coordinates:
$\textbf{x}=a e^{{\bf i}\,\vphi}+ be^{{\bf i}\,\psi}{\bf j}$.
Let us use  imaginary quaternions $\textbf{q} $ as stereogrphic
coordinates on $ {\bf H}_1$, and take symmetrical field
$\textbf{q}(\textbf{x}) $ consistent with Eq.~(\ref{f-so2}):
\begin{equation}\label{fq(x)}{
\textbf{q}(\textbf{x})=\textbf{x\,i}\,\bar{\textbf{x}}+\textbf{i}
=-\bar{\textbf{q}}, \ \textbf{f}(\textbf{x})=
-\frac{1+ \textbf{q} }{1- \textbf{q}}=
1-\frac{2}{1- \textbf{q}}. }
\end{equation}
It is easy to find the `center of quasi-soliton'
(1-submanifold, $S^1$)
\[ {
S^1=\textbf{f}^{-1}(-1)=\textbf{q}^{-1}(0)
=\{a=0,\ b=1\} =
\{\textbf{x}_0(\psi)=e^{{\bf i}\,\psi}{\bf j}\} }\]
and the `vector equipment' on this circle:
\[ { d \textbf{x}|_{\textbf{x}_0}\left.
=da\, e^{{\bf i}\,\vphi}+ (db +\textbf{i}\, d\psi)
e^{{\bf i}\,\psi}{\bf j}, \
\fr1 4  d \textbf{f}\right|_{\textbf{x}_0}=
{\bf i}db-{\bf k}\,
e^{{\bf i}\,(\vphi+\psi)} da\, ; }\]
\textbf{i}-vector all time looks along the radius $b$
(parallel translation along the circle $S^1$; this is a
`trivial`, or `flavor'-vector). Two others
('phase'-vectors) make $2\pi-$rotation along the circle.

In fact, the field (\ref{fq(x)}) has also symmetry
$SO_2\{1,4\}$, and this feature restricts possible
directions of `flavor'-vector (two `flavors' are
possible, $\pm$; the $P_2\{1,4\}-$symmetry
(this is the $\pi$-rotation of $x^1,x^4$) gives
the same effect).
The other interesting observation is that the
equipped circle
 can be located also at the stationary
points of $SO_2-$symmetry (this increases
the number of `flavors').

 \section{{\large Quasi-charges
 and their morphisms (in 5$D$, ie $m=4$) }}
If  $G\subset SO_4$,
the QC-group has two isomorphous parts,
  left and right: $\Pi(G)=\Pi_{(l)}(G)+\Pi_{(r)}(G)$.
  The  Table below describes quasi-charge groups
  for $
  G \subset G_0 = (O_3\times P_4)\cap SO_4\,
  $ ($P_4$ is spatial inversion,
  the 4-th coordinate is the
  extra dimension  of $G_0 $-symmetric
  expanding cosmological background).  

{\bf Table. } QC-groups  $\Pi_{(l)}(G)$ and their morphisms
to the preceding group; $G\subset G_0$.
\\[2mm] \hspace*{3\parindent}
\begin{tabular}{c|c|c}
 $G $ &$ \Pi_l(G)\to\Pi_l(G^*) $
&$ \mbox{`label'}$ \\ \hline\hline
$1 $&$ Z_2 $&  \\[1mm] \hline
$SO\{1,2\} $&$ Z_{(e)} \stackrel{e}{\to}Z_2 $&$ e $ \\[1mm] \hline
$SO\{1,2\}\times{}P\{3,4\} $&$ Z_{(\nu)}+Z_{(H)}
\stackrel{i,m2}{\to}Z_{(e)} $&$ \nu^0;\ H^0\to e+e \ \
$ \\[1mm] \hline\hline
$SO\{1,2\}\times{}P\{2,3\} $&$ Z_{(W)}
\stackrel{0}{\to}Z_{(e)} $&$ W\to e+\nu^{0}$ \\[1mm]
$SO\{1,2\}\times{}P\{2,4\} $&$ Z_{(Z)}
\stackrel{0}{\to}Z_{(e)} $&$ Z^0\to e+e$ \\[1mm]\hline
$SO\{1,2\}\times{}P\{3,4\}\times{} $&$ Z_{(\gamma)}
\stackrel{0}{\to}Z_{(H)} $&$ \gamma^0\to H^{0}+H^{0}$
 \\[0mm]
$ \times{}P\{2,3\} $&$ \ \ \ %
\stackrel{0}{\to} Z_{(W)}$ &$ \ \ \ \to W+W$
 \end{tabular}  \\[1mm]
`Quasi-particles', which symmetry includes $P_4$,
seem  to be true neutral (neutrinos, Higgs particles, photon).

  One can assume further that an hadron bag is a specific place
  where $G_0-$symmetry does not work, and the bag's symmetry
  is isomorphous to $O_4$.
  This assumption can lead to
  another classification of quasi-solitons (some doubling the
  above scheme), where self-dual and anti-self-dual one-parameter
  groups take place of $SO_2-$group.
  The total set of quasi-particle parameters
  (parameters of equipped 1-manifold (loop)  plus
  parameters of group) for (anti)self-dual groups,
 $G(4,2)\times RP^2$,
  is larger than the analogous set for
  groups $SO_2\subset G_0$,
  which is just $O_3\times G(3,1)=RP^2$ .
If the number of `flavor'-parameters (which are not degenerate and
have some preferable particular values;
this should be sensitive to discreet part of $G$ --
at least photons have the same flavor)
is the same as in the case of
`white' quasi-particles, the remaining parameters (degenerate, or
`phase') can give room for `color' (in addition to spin).
   So, perhaps one might think about `color neutrinos'
  (in the context of pomeron, and baryon spin puzzle),
`color W, Z, and Higgs' (another context -- $B$-mesons),
and so on.

Note that in this picture the very notion of quasi-particle depends on
the background symmetry (also to note: there are no
'quanta of torsion' per se). On the other hand, large clusters of
quasi-particles (matter) can disturb the background, and waves of such
small disturbances (with
wavelength  larger than the thickness $L$, perhaps)
can be generated as well (but these waves do not carry
(quasi)charges, that is, are {\sl not quantized}).

 \section{{\large Coexistence: phenomenological
 `quantum fields' on classical background }}
The non-linear, particle-like field
configurations with quasi-charges (quasi-particles) should be
very elongated along the extra-dimension
(all of the same size $L$),
while being small sized along usual dimensions, $\l\ll L$.
The motion of such a spaghetti-like quasi-particle should be
very complicated and stochastic due to `strong' imponderable
noise, such that different parts of spaghetti are coming
their own paths.
At the same time, quasi-particle can acquire
`its own' energy--momentum -- due to scattering of ponderable
waves (which  wave-vectors are almost tangent to usual 3$D$
(sub)space);
so, it seems that scattering
amplitudes\footnote{
These amplitudes can depend on additional vector-parameters
(`equipment vectors') relating to differential of field mapping  at a
`quasi-particle center' -- where quasi-charge density is
largest (if it has covariant sense).}
of those spaghetti's
parts which have the same 3$D$--coordinates can be summarized
providing an auxiliary, secondary field.

So, the imponderable waves provides stochasticity
(of motion of spaghetti's parts), while
the ponderable waves ensure superposition (with secondary fields).
Phenomenology of secondary fields could be of Lagrangian type,
with positive energy acquired by quasi-particles, -- that to
ensure the stability (of all the waveguide with its infill
-- with respect to quasi-particle production; the least action
principle has deep concerns with Lyapunov stability and is
deducible, in principle, from the path integral approach).

 \section{{\large `Plain' $R^2$ gravity on very thick
 brane \\ and change in the Newton's  Law of Gravitation}}
Let us start with 4d (from 5D) bi-Laplace equation with
a $\d$-source [as weak field, non-relativistic
(stationary) approximation
(it is assumed that `mass is possible')
 for R$^2$-gravity (\ref{tmunu})]
 and its solution ($R$ is 4d distance, radius):
\be\label{point} {
 \D^2\varphi = -\,\frac a {R^3}\,\d(R); \ \,
\varphi(R^2) = \frac a{\,8\,}\ln R^2 -\frac b {R^2} \
(+\,c\,, \mbox{ but $c$ does not matter}); }
\ee
the attracting force between two point masses is
$ F_{\rm point} = \frac a{4R} +\frac{2b}{R^3}$, $a,b$ should
be proportional to both masses.

Now let us suppose that all masses are distributed along
the extra dimension with a `universal function',
$\mu(p), \ \int\! \m(p)\,d p =1$. Then the attracting (gravitation)
force takes the next form [see (\ref{point}); $r$ is usual 3d distance]:
\be \label{mon} {
F(r)= \frac{d}{dr} 
\! \int\! \!\!\int^{{}^\infty}_{\!\!\!{}_{- \infty}}
\!\!\!\!\!
\varphi(r^2+(p-q)^2)\,\mu(p)\,\mu(q)\,dp\,d q
=\frac{a\,r}4\, V- b\, V\,' , \ \,
V(r)\, {=}\int\!\!\!\int\!\frac{\mu(p)\,\mu(q)\,dp\,d q}{r^2+(p-q)^2}.
 } \ee
(Note that $V(r)$ can be restored if $F(r)$ is measured.)

Taking
$\m_1(p)=\pi^{-1}/(1+p^2)$ (typical scale along
the extra dimension is taken as
unit, $L=1$; it seems that $L$ should be greater than ten AU),
one can find $rV_1(r)=1/(2+r)$ and
\[ {
F(r)=\frac{a}{8+4r}+\frac{2b(1+r)}{r^2(2+r)^2}; \ \,
\mbox{or (now $L\neq1$) }\,
 F(r)=\frac 1 {r^2} + \frac r{2L(2L+r)^2},
\mbox{ \ where $ a=b=2/L^2$}. } \]
Fig.\,1, curve\,(a) shows $\d F= F-1/r^2$ (deviation from the
Newton's Law; $a/b$ is chosen that
$\d F (0){=}0$);
two other curves, (b) \& (c), correspond to
$\mu_2=2\pi^{-1}\!/(1+p^2)^2, \ \mu_3=2\pi^{-1}p^2/(1+p^2)^2$
(also $\d F (0){=}0$; residues help to find $rV_2=(10+6r+r^2)/(2+r)^3$,
$rV_3=(2+2r+r^2)/(2+r)^3$).

We see that in principle this theory can explain galaxy rotation
curves,
$v^2(r)\,{\propto}\, rF\stackrel{r\to\infty}{\longrightarrow}\,$const,
 without need for Dark Matter (or MOND \cite{mond};
about rotation curves and DM see
\cite{peskin}; they are looking for DM in Solar system too,
\cite{iorio}).

Q: Can the `coherence of mass' along the extra dimension be
disturbed ? (the flyby anomaly,
the Pioneer anomaly \cite{LPD}); can $\mu(p)$
be negative  in some domains of $p$ ?
\vspace{0.1mm}
\begin{figure}
  \centering
\includegraphics[bb=25 10 290 230,height=100mm]{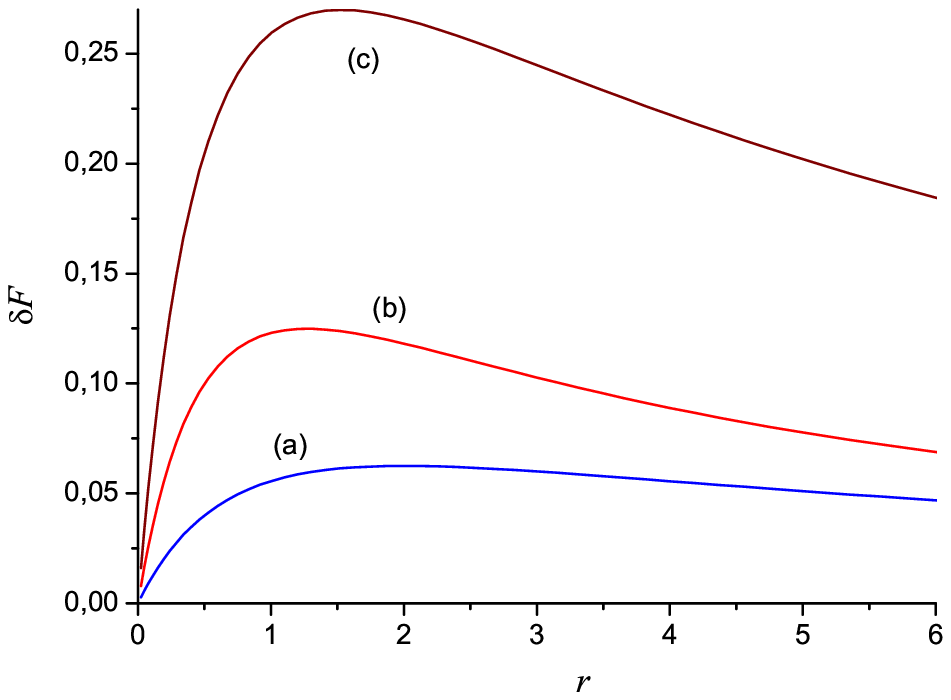}
 ~\\ 
\centerline{ Fig.\,1. \ Deviation $\d F= F-1/r^2$
  for different $\mu(p)$, see  Eq.~(\ref{mon}) and text below.}
\end{figure}
\ ~
 \section{{\large How to register `powerless' waves }}
This section is added perhaps for some funny recreation (or
still not ? who knows). We have learnt that $S$-waves do not
carry momentum and angular momentum, so they can not perform any
work or spin flip.

But let us conceive that these waves can effect
a flip-flop of two neighbor spins. So,
a `detector' could be a media with two sorts of spins,
A and B. Let $s^A=s^B=1/2$ but
$g^A\neq g^B$, and let the initial state
is prepared as follows:
$\{{<}s^A_z{>},{<}s^B_z{>}\}(0)=\{1/2,-1/2\} $.
Then the process of spin relaxation starts; turning on
appropriate magnetic field $H_z$ (and alternating
fields of proper frequencies) one can measure the
detector's state and find the time of spin relaxation.

The next step. Skilled experimenters  
try to generate $S$-waves
and to register an effect of these waves on spin
relaxation. The generation of intense `coherent'
$S$-waves could be proceeded perhaps with a similar
spin system subjected to alternating polarization.
 \section{{\large Single photon experiment
 (that to feel huge extra dimension), and Conclusion }}
Today, many laboratories have sources of single
(heralded) photons, or entangled bi-photons
(say, for Bell-type experiments \cite{weihs});
some students can perform laboratory works
with single photons, having convinced on their own experience
that light is quantized
(the Grangier experiment)\cite{sites}.

It is being suggested a minor modification
of the single (polarized) photon interference experiment,
say, in a
Mach-Zehnder fiber interferometer with `long' (the fibers may be
rolled) enough arms.
The only new element  is a fast-acting
shutter placed at the beginning of (one of) the interferometer's
arms (the closing-opening time of the shutter should
be smaller than the flight time in the arms).
For example, a fast electro-optical modulator
in combination with polarizer (or a number of
such combinations) can be used with polarized photons.

Both Quantum mechanics (no particle's ontology) and Bohmian
mechanics (wave-particle double ontology)\cite{nik}
 exclude any change in
the interference figure as a result of separating activity of
such a fast shutter (while the photon's `halves' are making their
ways to the place of a meeting).
However, if a photon has non-local spaghetti-like ontology
(along the extra dimension) and fragments of this
spaghetti are moving along both arms at once, then the
 shutter should tear up this spaghetti
 (mainly without photon absorption), tear out
 its fragments (which will dissolve in `zero-point
oscillations').
 Hence, if the absorption factor of the shutter
 (the extinction ratio of polarizer) is large enough,
the 50/50-proportion (between the photon's amplitudes
in the arms) will be changed and
\emph{a significant decrease of the interference
visibility should be observed}.

 QM is everywhere (where we can see, of course),
 and, so, non-linear 5$D$-field
fluctuations, looking like spaghetti-anti-spaghetti loops,
should exist everywhere. (This omnipresence can be
related to the universality of `low-level heat death',
restricted by
the presence of topological quasi-solitons -- some as
 the 2$D$ computer experiment by Fermi, Pasta, and Ulam,
where the process of thermalization was restricted by
the existence of solitons. See also the sections 5--8
(and \cite{tc})
for arguments in favor of phenomenological (quantized)
`secondary fields' accounting for topological
(quasi)charges and obeying superposition, path integral
and so on.)

AP, at least at the level of its symmetry,
seems to be able to cure the
gap between the two branches of physics -- General Relativity
(with coordinate diffeomorphisms) and Quantum Mechanics
(with Lorentz invariance).\footnote{
 Rovelli writes\cite{rov}:
{\it In spite of their empirical success, GR and QM offer
a schizophrenic and confused understanding
of the physical world}.}
Most people give all the rights of fundamentality to
quanta, and so, they are trying to quantize gravity, and the
very space-time (probing loops, and strings, and branes;
see also the warning polemic by Schroer \cite{schroer}).
The other possibility is that quanta have the specific
phenomenological origin relating to topological (quasi)charges.


\end{document}